\documentclass[pra,twocolumn,showpacs]{revtex4-1}
\usepackage{amsmath,amssymb,bm,graphicx,hyperref}

\newcommand\+{\dagger}
\newcommand\<{\langle}
\renewcommand\>{\rangle}
\newcommand\up{\uparrow}
\newcommand\down{\downarrow}
\renewcommand\k{{\bm{k}}}
\renewcommand\r{{\bm{r}}}
\newcommand\eps{\epsilon}
\newcommand\ek{\eps_\k}
\newcommand\kB{k_\mathrm{B}}
\newcommand\kF{k_\mathrm{F}}
\newcommand\eF{\eps_\mathrm{F}}
\newcommand\TF{T_\mathrm{F}}
\newcommand\TK{T_\mathrm{K}}

\renewcommand\Im{\mathrm{Im}}
\newcommand\Tr{\mathrm{Tr}}
\newcommand\G{\mathcal{G}}
\newcommand\R{\mathcal{R}}
\newcommand\T{\mathcal{T}}

\begin{document}

\title{Transport measurement of the orbital Kondo effect with ultracold atoms}

\author{Yusuke Nishida}
\affiliation{Department of Physics, Tokyo Institute of Technology,
Ookayama, Meguro, Tokyo 152-8551, Japan}

\date{August 2015}

\begin{abstract}
 The Kondo effect in condensed-matter systems manifests itself most
 sharply in their transport measurements.  Here we propose an analogous
 transport signature of the orbital Kondo effect realized with ultracold
 atoms.  Our system consists of imbalanced Fermi seas of two components
 of fermions and an impurity atom of different species which is confined
 by an isotropic potential.  We first apply a $\pi/2$ pulse to transform
 two components of fermions into two superposition states.  Their
 interactions with the impurity atom then cause a ``transport'' of
 fermions from majority to minority superposition states, whose numbers
 can be measured after applying another $3\pi/2$ pulse.  In particular,
 when the interaction of one component of fermions with the impurity
 atom is tuned close to a confinement-induced $p$-wave or higher
 partial-wave resonance, the resulting conductance is shown to exhibit
 the Kondo signature, i.e., universal logarithmic growth by lowering the
 temperature.  The proposed transport measurement will thus provide a
 clear evidence of the orbital Kondo effect accessible in ultracold atom
 experiments and pave the way for developing new insights into Kondo
 physics.
\end{abstract}


\maketitle

\section{Introduction}
The Anderson impurity model is one of the most important and fundamental
model Hamiltonians in condensed-matter physics~\cite{Anderson:1961}.  It
was originally invented to study localized magnetic impurities in
metallic environments and there exist three distinct parameter regimes
called empty orbital regime, mixed valence regime, and local moment
regime~\cite{Hewson:1993}.  Of particular interest is the local moment
regime, where the electrical resistivity grows logarithmically toward
the low temperature as a consequence of the celebrated Kondo
effect~\cite{Kondo:1964,Schrieffer:1966}.  By extending the Anderson
impurity model, orbital degeneracy can also be
incorporated~\cite{Coqblin:1969} and periodic arrangement of magnetic
impurities is considered to be relevant to heavy fermion
physics~\cite{Coleman:2007}.

Further application of the Anderson impurity model is possible by
introducing additional degrees of freedom corresponding to left and
right leads to study tunneling of electrons through a quantum dot both
in and out of equilibrium~\cite{Meir:1991,Meir:1993}.  In fact, one
degree of freedom can be decoupled by a canonical transformation and the
other interacting degree of freedom turns out to be described by the
original Anderson impurity model~\cite{Glazman:1988}.  Therefore, the
Kondo effect emerges again in quantum dot systems but now as the
logarithmic growth of the electrical conductance by lowering the
temperature~\cite{Glazman:1988,Ng:1988}, where a number of beautiful
observations of the Kondo effect have been
made~\cite{Grobis:2007,Wiel:2010}.

The purpose of this Rapid Communication is to show that all the rich
physics associated with the Anderson impurity model can be simulated
with ultracold atoms by employing standard techniques such as
magnetic-field-induced Feshbach resonances, species-selective optical
lattices, and laser couplings of atomic hyperfine
states~\cite{Lewenstein:2012}.  In particular, we place special emphasis
on the transport measurement of the Kondo effect analogous to quantum
dot experiments, which should be of great importance in ultracold atom
experiments because transport is usually difficult to study with the
exception of recent progress made in
Refs.~\cite{Brantut:2012,Stadler:2012,Krinner:2013,Brantut:2013,Krinner:2015,Lee:2015,Husmann:2015}.
Future realization of the Kondo effect and its transport measurement
with ultracold atoms will pave the way for developing new insights into
yet unresolved aspects of Kondo physics such as the formation and
dynamics of the Kondo screening cloud~\cite{Kouwenhoven:2001} and the
quantum criticality in heavy fermion systems~\cite{Coleman:2007}.

\section{Setup and measurement protocol}
Our study is based on the simple and versatile scheme to realize the
orbital Kondo effect with ultracold atoms~\cite{Nishida:2013} (see
Refs.~\cite{Recati:2002,Falco:2004,Duan:2004,Paredes:2005,Gorshkov:2010,Lal:2010,Bauer:2013,Kuzmenko:2015,Sundar:2015,Nakagawa:2015,Patton:2015}
for other proposals).  The system consists of a Fermi sea of
spin-polarized $\up$ fermions of species $A$ interacting with a spinless
impurity atom of different species $B$ which is loaded into a ground
state of an isotropic potential.  By tuning the interspecies attraction
with an $s$-wave Feshbach resonance, the impurity atom and a
spin-polarized fermion can form a bound molecule that occupies a
degenerate orbital of the confinement potential with orbital angular
momentum $\ell\geq1$~\cite{Nishida:2010}.  In particular, when the total
energy of the bound molecule coincides with the scattering threshold of
the $A_\up$ and $B$ atoms, an $\ell$th partial-wave resonance is induced
and low-energy physics in its vicinity is described by a two-channel
Hamiltonian:
\begin{align}\label{eq:H_up}
 H_\up &= \int\!\frac{d\k}{(2\pi)^3}\,\ek\,\psi_{A\up}^\+(\k)\psi_{A\up}(\k)
 + \sum_{m=-\ell}^\ell\delta_m\phi_m^\+\phi_m \notag\\
 & + \sum_{m=-\ell}^\ell\int\!\frac{d\k}{(2\pi)^3}
 \left[V_\ell^m(\k)\psi_{A\up}^\+(\k)\psi_B^\+\phi_m + \text{H.c.}\right].
\end{align}
Here $\psi_{A\up}^\+(\k)$ creates an $A_\up$ atom with energy
$\ek=\hbar^2\k^2/(2M)$, while $\psi_B^\+$ creates the impurity $B$ atom
in the ground state of the confinement potential whose energy is chosen
to be zero.  The bound molecule is created by $\phi_m^\+$ in one of the
degenerate orbitals labeled by the magnetic quantum number
$|m|\leq\ell$.  Its coupling to the $A_\up$ and $B$ atoms is assumed to
have the harmonic form of
\begin{align}\label{eq:coupling}
 V_\ell^m(\k) = v_m|\k|^\ell Y_\ell^m(\hat\k)
 \exp\!\left[-\frac{\k^2}{2\Lambda^2}\right]
\end{align}
with the wave-number cutoff $\Lambda$ set by an inverse characteristic
extent of the confined $B$ atom.  Because only one $B$ atom is confined,
the particle number operators of the localized $B$ atom and bound
molecule are constrained by
\begin{align}\label{eq:constraint}
 N_B = \psi_B^\+\psi_B + \sum_{m=-\ell}^\ell\phi_m^\+\phi_m = 1.
\end{align}
When the rotational symmetry is exact, we have an equal detuning
$\delta_m$ and coupling $v_m$ for all $m$ and thus the degeneracy is
($2\ell+1$)-fold, while we shall develop a general formulation so that
it is also applicable to study the effect of symmetry breaking later.

Interestingly, the low-energy effective Hamiltonian (\ref{eq:H_up})
naturally realizable with ultracold atoms is nothing but the
infinite-$U$ Anderson impurity model in the slave-particle
representation~\cite{Barnes:1976,Coleman:1984} with its fictitious
degrees of freedom corresponding to our real atom and molecule as in
Eq.~(\ref{eq:constraint}).  The empty orbital, mixed valence, and local
moment regimes of the original Anderson impurity model are thus
translated into atomic regime ($\<\psi_B^\+\psi_B\>\simeq1$), resonant
regime ($0\lesssim\<\psi_B^\+\psi_B\>\lesssim1$), and molecular regime
($\<\psi_B^\+\psi_B\>\simeq0$), respectively, in the language of
ultracold atoms.  In particular, the orbital Kondo effect emergent in
the molecular limit was elaborated for $\ell=1$ in
Ref.~\cite{Nishida:2013}, while the analysis therein can be
straightforwardly generalized for an arbitrary $\ell$ to find that the
Kondo temperature in the SU($2\ell+1$) symmetric case has a universal
leading exponent given by
\begin{align}\label{eq:T_kondo}
 \TK \propto \TF\exp\!\left[-\frac{\pi}{(2\ell+1)a_\ell\kF^{2\ell+1}}\right]
\end{align}
with $a_\ell\ll\kF^{-2\ell-1}$ being the $\ell$th partial-wave
scattering length.  Because the Kondo effect in condensed-matter systems
manifests itself most sharply in their transport measurements, it is
highly desired although challenging to establish an analogous transport
signature of the orbital Kondo effect in ultracold atom experiments.

In what follows, we indeed show that the conductance measurement of the
Kondo effect in quantum dot experiments can be equivalently performed
with ultracold atoms by adopting the idea from Ref.~\cite{Knap:2012}.
To this end, we introduce another spin $\down$ component of fermionic
species $A$,
\begin{align}\label{eq:H_down}
 H_\down = \int\!\frac{d\k}{(2\pi)^3}\,\ek\,\psi_{A\down}^\+(\k)\psi_{A\down}(\k),
\end{align}
as well as the intercomponent coupling driven by a resonant laser field
with the Rabi frequency $\Omega$,
\begin{align}\label{eq:H_up-down}
 H_{\up\down} = i\frac{\hbar\Omega}{2}\int\!\frac{d\k}{(2\pi)^3}
 \left[\psi_{A\up}^\+(\k)\psi_{A\down}(\k) - \psi_{A\down}^\+(\k)\psi_{A\up}(\k)\right],
\end{align}
which are expressed in the rotating frame.  It is legitimate to assume
that interactions of $A_\down$ atoms with $A_\up$ atoms and with the
impurity $B$ atom are both negligible in the dilute limit because they
are generally off-resonance when the interaction of $A_\up$ atoms with
the impurity $B$ atom is tuned close to a confinement-induced resonance.
It is also important that the confinement-induced resonance can be
turned on and off without changing the magnetic field but by controlling
the potential strength acting on the impurity $B$
atom~\cite{Nishida:2010}.  With all these setups, we are ready to
propose a simple conductance measurement with ultracold atoms consisting
of the following three protocols.

(i) Preparation: We first stay away from any confinement-induced
resonances so that both $A_\up$ and $A_\down$ atoms negligibly interact
with the impurity $B$ atom.  After introducing $N_-$ and $N_+$ numbers
of $A_\up$ and $A_\down$ atoms, we apply the intercomponent coupling
(\ref{eq:H_up-down}) for a duration of $\pi/(2\Omega)$, which transforms
the two spin components into two superposition states according to
$|A_\up\>\to|A_-\>\equiv(|A_\up\>-|A_\down\>)/\sqrt2$ and
$|A_\down\>\to|A_+\>\equiv(|A_\up\>+|A_\down\>)/\sqrt2$.  Consequently,
$N_-$ and $N_+$ numbers of $A_-$ and $A_+$ atoms are prepared at a
common temperature $T$ and their corresponding chemical potentials are
denoted by $\mu_-\equiv\mu-\Delta\mu/2$ and
$\mu_+\equiv\mu+\Delta\mu/2$, respectively.

(ii) Transport: We now turn on the confinement-induced $\ell$th
partial-wave resonance of $A_\up$ atoms with the impurity $B$ atom.
Because $A$ atoms are prepared on the $\pm$ basis, it is appropriate to
express the total Hamiltonian composed of Eqs.~(\ref{eq:H_up}) and
(\ref{eq:H_down}) in terms of the corresponding creation operators
$\psi_{A\pm}^\+(\k)\equiv[\psi_{A\up}^\+(\k)\pm\psi_{A\down}^\+(\k)]/\sqrt2$
as
\begin{align}\label{eq:transport}
 & H_\up + H_\down = \sum_{\sigma=\pm}\int\!\frac{d\k}{(2\pi)^3}\,\ek\,
 \psi_{A\sigma}^\+(\k)\psi_{A\sigma}(\k)
 + \sum_{m=-\ell}^\ell\delta_m\phi_m^\+\phi_m \notag\\
 & + \sum_{\sigma=\pm}\sum_{m=-\ell}^\ell\int\!\frac{d\k}{(2\pi)^3}
 \left[\frac{V_\ell^m(\k)}{\sqrt2}\psi_{A\sigma}^\+(\k)\psi_B^\+\phi_m
 + \text{H.c.}\right],
\end{align}
where $A_-$ and $A_+$ atoms equally interact with the impurity $B$
atom.  Remarkably, the resulting Hamiltonian is identical to the
Anderson impurity model extended to study tunneling of electrons through
a quantum dot~\cite{Meir:1991,Meir:1993} with roles of left and right
leads played by our two superposition states.  The chemical potential
imbalance $\Delta\mu\neq0$ thus causes a transport of fermions from
majority to minority superposition states through scatterings with the
impurity $B$ atom~\cite{transport}.  After a period of $\Delta t$, the
numbers of $A_-$ and $A_+$ atoms change into $N_-+I\Delta t$ and
$N_+-I\Delta t$ with $I$ being the steady-state current.

(iii) Measurement: We then turn off the confinement-induced resonance to
terminate the transport.  By applying the intercomponent coupling
(\ref{eq:H_up-down}) again but for a duration of $3\pi/(2\Omega)$, the
two superposition states are transformed back into the original two spin
components as $|A_-\>\to-|A_\up\>$ and $|A_+\>\to-|A_\down\>$ whose
numbers are now $N_-+I\Delta t$ and $N_+-I\Delta t$, respectively.
Finally, from the measured numbers of $A_\up$ and $A_\down$ atoms
compared to their initial values, the transported fermion number
$I\Delta t$ can be extracted to determine the conductance $I/\Delta\mu$
of the transport Hamiltonian (\ref{eq:transport}).  This constitutes the
ultracold atom equivalent of the conductance measurement in quantum dot
systems.

\section{Conductance and the Kondo effect}
In order to provide quantitative guides on how the orbital Kondo effect
emerges in the proposed conductance measurement with ultracold atoms, we
study in detail the linear conductance
$G=\lim_{\Delta\mu\to0}I/\Delta\mu$ of the transport Hamiltonian
(\ref{eq:transport}).  With the aid of the Meir-Wingreen formula for the
steady-state current $I=(dN_-/dt-dN_+/dt)/2$~\cite{Meir:1992}, the
linear conductance is given by
\begin{align}\label{eq:conductance}
 G = \frac1\hbar\sum_{m=-\ell}^\ell\int\!\frac{d\k}{(2\pi)^3}
 \frac{|V_\ell^m(\k)|^2}{2}f'_T(\ek-\mu)\,\Im\,\G_m^R(\ek-\mu),
\end{align}
where $f_T(z)=1/(e^{\beta z}+1)$ is the Fermi-Dirac distribution
function at an inverse temperature $\beta=1/(\kB T)$ and
\begin{align}\label{eq:retarded}
 \G_m^R(\eps) = -\frac{i}{\hbar}\int_0^\infty\!dt\,e^{i\eps t/\hbar}
 \<\{\psi_B^\+(t)\phi_m(t),\phi_m^\+(0)\psi_B(0)\}\>
\end{align}
is the retarded Green's function with the expectation value taken with
respect to the equilibrium state at $\Delta\mu=0$.  Therefore, $A_-$ and
$A_+$ atoms now have the equal chemical potential $\mu_\pm=\mu$ and thus
it is advantageous to express the transport Hamiltonian
(\ref{eq:transport}) on the original spin basis so that it is decoupled
into Eqs.~(\ref{eq:H_up}) and (\ref{eq:H_down})~\cite{decoupling}.
Because $A_\down$ atoms do not interact with the impurity $B$ atom, the
expectation value is simply evaluated as
$\<\cdots\>=\Tr[e^{-\beta(H_\up-\mu N_\up)}\cdots]/Z$ along with the
constraint (\ref{eq:constraint}), where
\begin{align}
 N_\up = \int\!\frac{d\k}{(2\pi)^3}\psi_{A\up}^\+(\k)\psi_{A\up}(\k)
 + \sum_{m=-\ell}^\ell\phi_m^\+\phi_m
\end{align}
is the particle number operator of $A_\up$ atoms.  While the equilibrium
Green's function (\ref{eq:retarded}) can be computed by means of various
methods~\cite{Georges:1996}, we here employ the so-called noncrossing
approximation~\cite{Bickers:1987,Hewson:1993}, which is known to be
reliable for large degeneracy and not too low temperature, and thus
adequate for our purpose.

\begin{figure}[t]
 \includegraphics[width=\columnwidth]{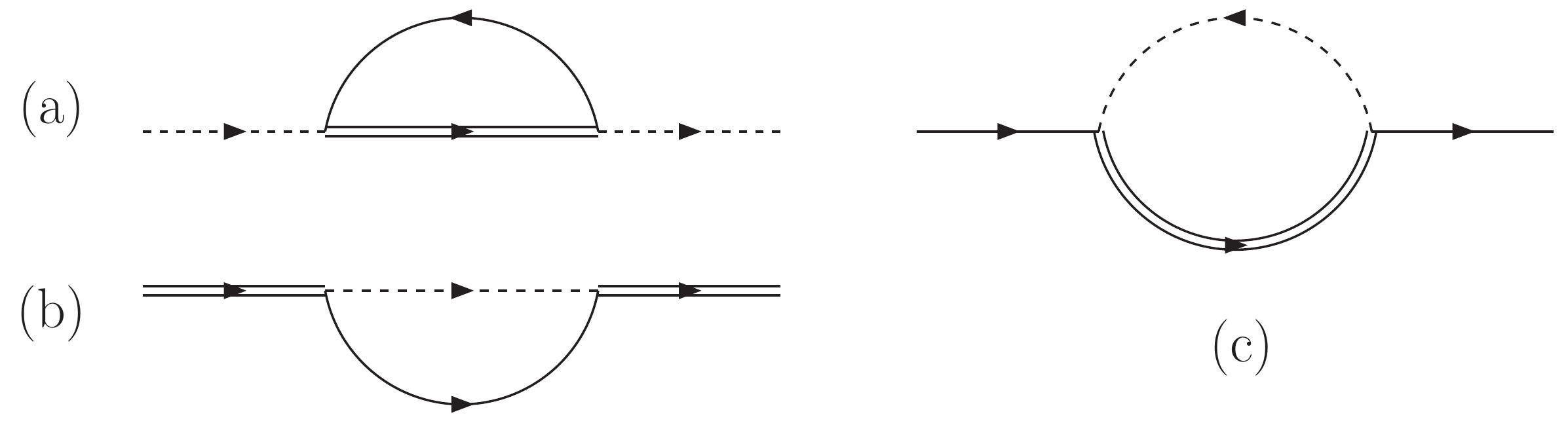}
 \caption{Noncrossing approximation for self-energies of the localized
 (a) $B$ atom [Eq.~(\ref{eq:self-energy_B})] and (b) bound molecule
 [Eq.~(\ref{eq:self-energy_m})] as well as (c) the Matsubara Green's
 function corresponding to Eq.~(\ref{eq:retarded}).  Solid, dashed, and
 doubled lines represent the propagators of $A_\up$ atom, $B$ atom, and
 bound molecule, respectively.  \label{fig:noncrossing}}
\end{figure}

The propagators of the localized $B$ atom and bound molecule are
sometimes called resolvents and denoted by
$\R_B(z)=[z-\Sigma_B(z)]^{-1}$ and
$\R_m(z)=[z+\mu-\delta_m-\Sigma_m(z)]^{-1}$, respectively.  Their
self-energies within the noncrossing approximation are depicted in
Fig.~\ref{fig:noncrossing} and determined self-consistently according to
\begin{align}\label{eq:self-energy_B}
 \Sigma_B(z) = \sum_{\ell=-m}^m\int\!\frac{d\k}{(2\pi)^3}\,
 |V_\ell^m(\k)|^2\,f_T(\ek-\mu)\,\R_m(z+\ek-\mu)
\end{align}
and
\begin{align}\label{eq:self-energy_m}
 \Sigma_m(z) = \int\!\frac{d\k}{(2\pi)^3}\,
 |V_\ell^m(\k)|^2\,f_T(\mu-\ek)\,\R_B(z-\ek+\mu).
\end{align}
In terms of the corresponding spectral densities of
$\rho_B(\eps)=-(1/\pi)\,\Im\,\R_B(\eps+i0^+)$ and
$\rho_m(\eps)=-(1/\pi)\,\Im\,\R_m(\eps+i0^+)$, the imaginary part of the
retarded Green's function (\ref{eq:retarded}) is expressed as
\begin{align}\label{eq:imaginary}
 -\frac1\pi\,\Im\,\G_m^R(\eps)
 = \frac{1+e^{-\beta\eps}}{Z_B}\int_{-\infty}^\infty\!dz\,
 e^{-\beta z}\rho_B(z)\rho_m(z+\eps)
\end{align}
with
\begin{align}
 Z_B = \int_{-\infty}^\infty\!dz\,e^{-\beta z}
 \left[\rho_B(z)+\sum_{m=-\ell}^\ell\rho_m(z)\right]
\end{align}
being the impurity partition function~\cite{Bickers:1987,Hewson:1993}.
The substitution of Eq.~(\ref{eq:imaginary}) into
Eq.~(\ref{eq:conductance}) now allows us to compute the linear
conductance numerically for a given set of parameters.

Besides the chemical potential $\mu$ and temperature $T$ of $A_\up$
atoms, the linear conductance (\ref{eq:conductance}) depends on the
detuning $\delta_m$ and coupling $v_m$ as well as the wave-number cutoff
$\Lambda$ through $V_\ell^m(\k)$ defined in Eq.~(\ref{eq:coupling}).  In
order to make contact with ultracold atom experiments, the bare
parameters $\delta_m$ and $v_m$ should be expressed in terms of physical
parameters such as the scattering length $a_m$ and the resonance range
$r_m$ characterizing low-energy scatterings in the $\ell$th partial-wave
channel.  They can be related by matching the two-body scattering
$\T$ matrix in the vacuum computed from the two-channel Hamiltonian
(\ref{eq:H_up}) with the standard form of
\begin{align}\label{eq:T-matrix}
 \T_\ell(k) = \frac{8\pi^2\hbar^2}{M}\sum_{m=-\ell}^\ell
 \frac{k^{2\ell}\,Y_\ell^m(\hat\k_\mathrm{out})\bar{Y}_\ell^m(\hat\k_\mathrm{in})}
 {ik^{2\ell+1}+1/a_m+r_mk^2+O(k^4)},
\end{align}
where we find
\begin{align}
 \frac{1}{a_m} = -\frac{8\pi^2\hbar^2\delta_m}{Mv_m^2}
 + \frac{\Gamma(\ell+\frac12)\Lambda^{2\ell+1}}{\pi}
\end{align}
and
\begin{align}
 r_m = \frac{1}{a_m\Lambda^2} + \frac{4\pi^2\hbar^4}{M^2v_m^2}
 + \frac{\Gamma(\ell-\frac12)\Lambda^{2\ell-1}}{\pi}.
\end{align}
In the vicinity of the confinement-induced resonance
$1/a_m\ll\Lambda^{2\ell+1}$, the two-body scattering $\T$ matrix
(\ref{eq:T-matrix}) has a pole in terms of the scattering energy
$\eps=\hbar^2k^2/(2M)$ at $\eps_m=-\hbar^2/(2Ma_mr_m)$, which is the
physical molecular energy and tunable in ultracold atom experiments.
These dimensionful quantities are customarily normalized with respect to
the Fermi wave number $\kF$ defined through the particle number density
of $A_\up$ atoms as $\kF^3/(6\pi^2)=\int\!d\k/(2\pi)^3f_T(\ek-\mu)$.
Consequently, the linear conductance (\ref{eq:conductance}) is
parameterized by $\eps_m/\eF$, $r_m/\kF^{2\ell-1}$, $\Lambda/\kF$, and
$T/\TF$ with $\eF=\hbar^2\kF^2/(2M)$ and $\TF=\eF/\kB$ being the Fermi
energy and temperature, respectively.

\begin{figure}[t]
 \includegraphics[width=0.92\columnwidth]{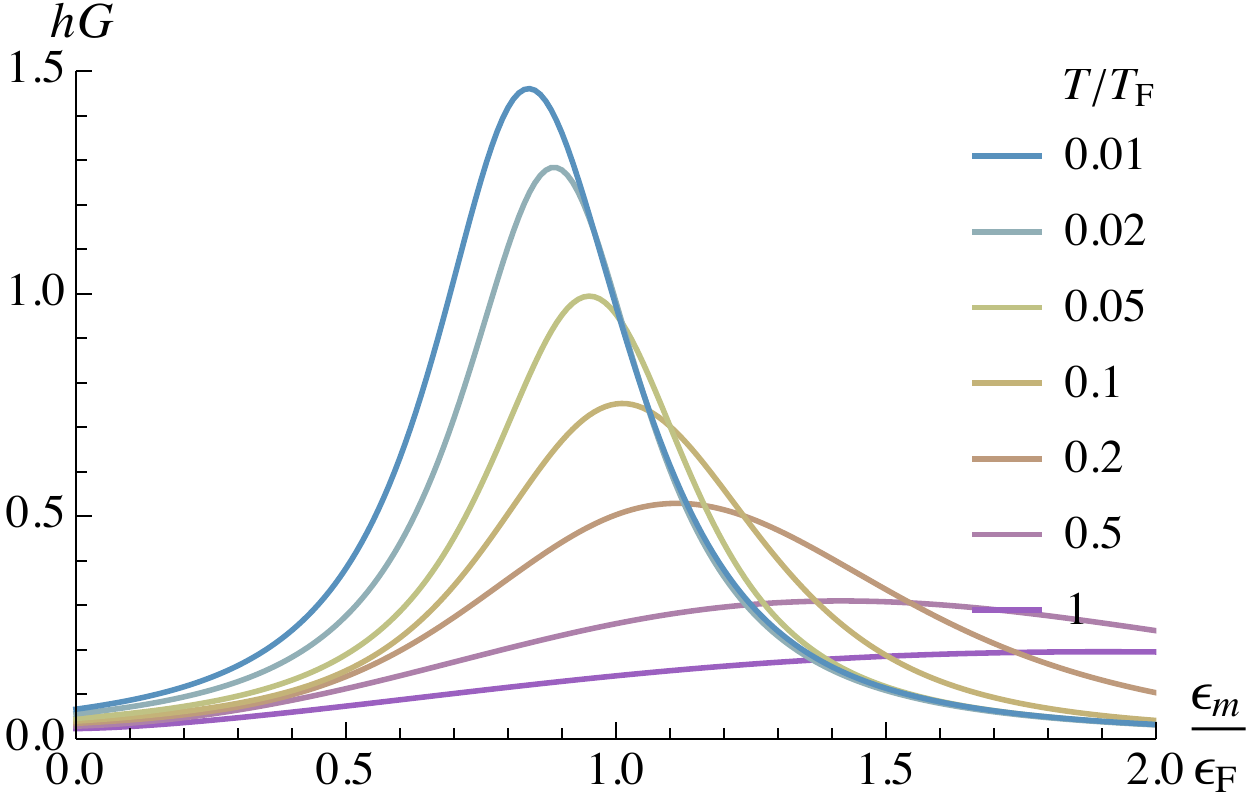}\bigskip\\
 \includegraphics[width=0.92\columnwidth]{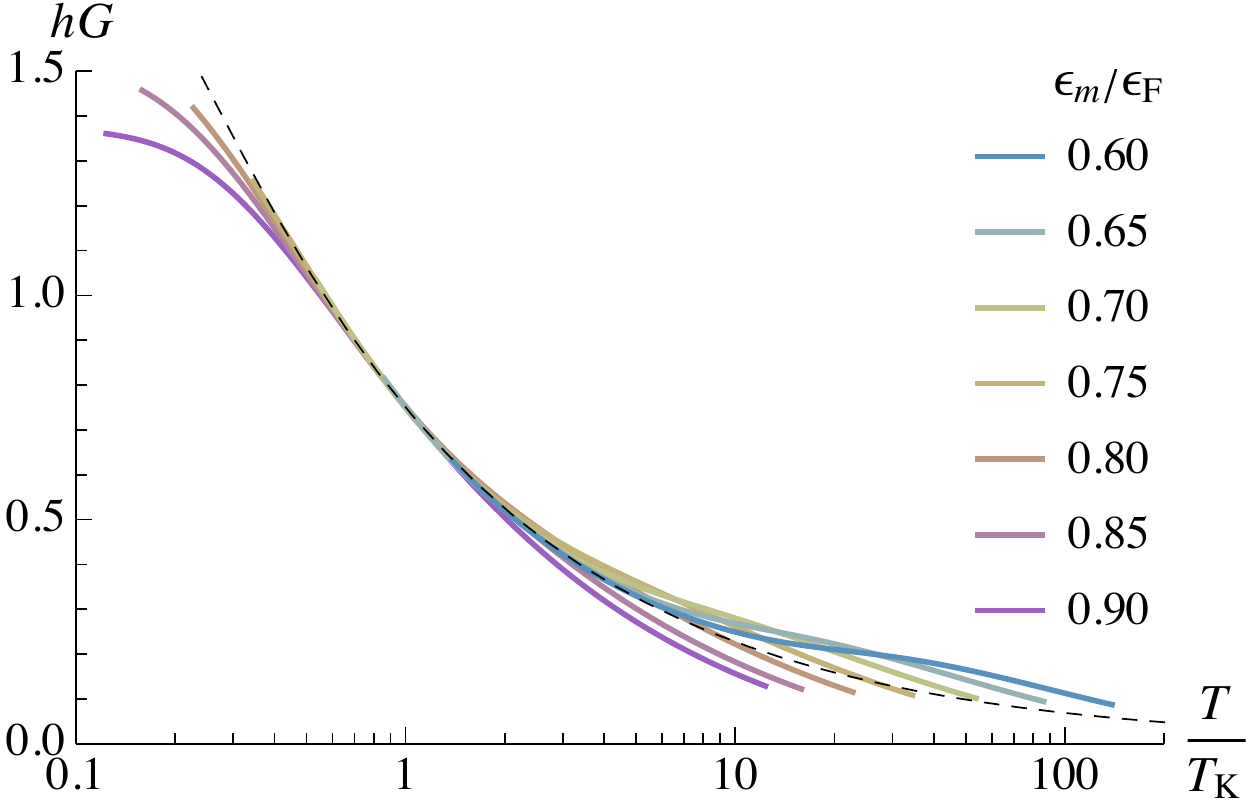}
 \caption{Linear conductance $G$ in the SU(3) symmetric case with
 $\ell=1$, $\Lambda=10\kF$, and $r_m=\Lambda$ as a function of the
 threefold degenerate molecular energy $\eps_m=-\hbar^2/(2Ma_mr_m)$ for
 selected $T/\TF=0.01$, 0.02, 0.05, 0.1, 0.2, 0.5, and 1 from highest to
 lowest curves (upper panel) and as a function of the temperature $T$
 for selected $\eps_m/\eF=0.6$, 0.65, 0.7, 0.75, 0.8, 0.85, and 0.9 from
 rightmost to leftmost curves (lower panel).  All curves in the lower
 panel are plotted in the same temperature range of
 $0.01\TF\leq T\leq\TF$ but normalized individually by $\TK$ for each
 $\eps_m/\eF$ so as to best fit to a common empirical form
 (\ref{eq:empirical}) of $hG(T)\approx(9/4)/[1+67.(T/\TK)^2]^{0.26}$
 indicated by the dashed curve.  \label{fig:SU3-R1}}
\end{figure}

While the formulation developed so far is general, we now focus on the
confinement-induced $p$-wave resonance in a dilute system with $\ell=1$
and $\Lambda=10\kF$.  Figure~\ref{fig:SU3-R1} shows the computed linear
conductance $G$ in units of the Planck constant $h=2\pi\hbar$ in the
SU(3) symmetric case with the resonance range $r_m=\Lambda$ chosen
corresponding to the case where the confined $B$ atom is lighter than
$A$ atoms~\cite{Nishida:2010}.  In the upper panel, $hG$ is plotted as a
function of the threefold degenerate molecular energy
$0\leq\eps_m/\eF\leq2$ for selected temperatures $T/\TF=0.01$, 0.02,
0.05, 0.1, 0.2, 0.5, and 1 accessible in ultracold atom experiments.
When the molecular energy is above the Fermi energy
($\eps_m\gtrsim\eF$), the conductance decreases toward the low
temperature until it eventually reaches a single Lorentzian curve with
its width set by the molecular lifetime.  On the other hand, when the
molecular energy is below the Fermi energy ($\eps_m\lesssim\eF$), the
conductance exhibits a strikingly distinct behavior, i.e., logarithmic
growth by lowering the temperature, which is the hallmark of the Kondo
effect.  Actually, the most remarkable feature of the Kondo effect is
the fact that the conductance turns out to be the universal function of
the temperature characterized by a single quantity $\TK$ in which all
microscopic details are encoded~\cite{Costi:1994}.  Such universality is
demonstrated in the lower panel of Fig.~\ref{fig:SU3-R1} where $hG$ for
selected molecular energies $\eps_m/\eF=0.6$, 0.65, 0.7, 0.75, 0.8,
0.85, and 0.9 are plotted in the same temperature range of
$0.01\TF\leq T\leq\TF$ but normalized individually by
$\TK\approx0.0072\TF$, $0.012\TF$, $0.019\TF$, $0.029\TF$, $0.044\TF$,
$0.063\TF$, and $0.081\TF$ so that all curves come together to make up a
single universal curve as much as possible.  The obtained universal
function is found to be well described by the empirical
form~\cite{Goldhaber-Gordon:1998} of
\begin{align}\label{eq:empirical}
 G(T) = \frac{G_0}{[1+(3^{1/\gamma}-1)(T/\TK)^2]^\gamma},
\end{align}
where $G_0=(2\ell+1)\sin^2[\pi/(2\ell+1)]/h$ is the zero temperature
conductance for the SU($2\ell+1$) Kondo effect in our
realization~\cite{Carmi:2011}, the Kondo temperature is defined by
$G(\TK)=G_0/3$, and the exponent $\gamma\approx0.26$ is chosen so as to
achieve the best fit to the numerical data.  While the Kondo temperature
decreases exponentially toward the molecular limit
$\eps_m/\eF\to-\infty$ as in Eq.~(\ref{eq:T_kondo}), we find a
reasonable range of molecular energy $0.5\lesssim\eps_m/\eF\lesssim1$
where the universal logarithmic growth of the conductance is observable
in ultracold atom experiments.

\begin{figure}[t]
 \includegraphics[width=0.92\columnwidth]{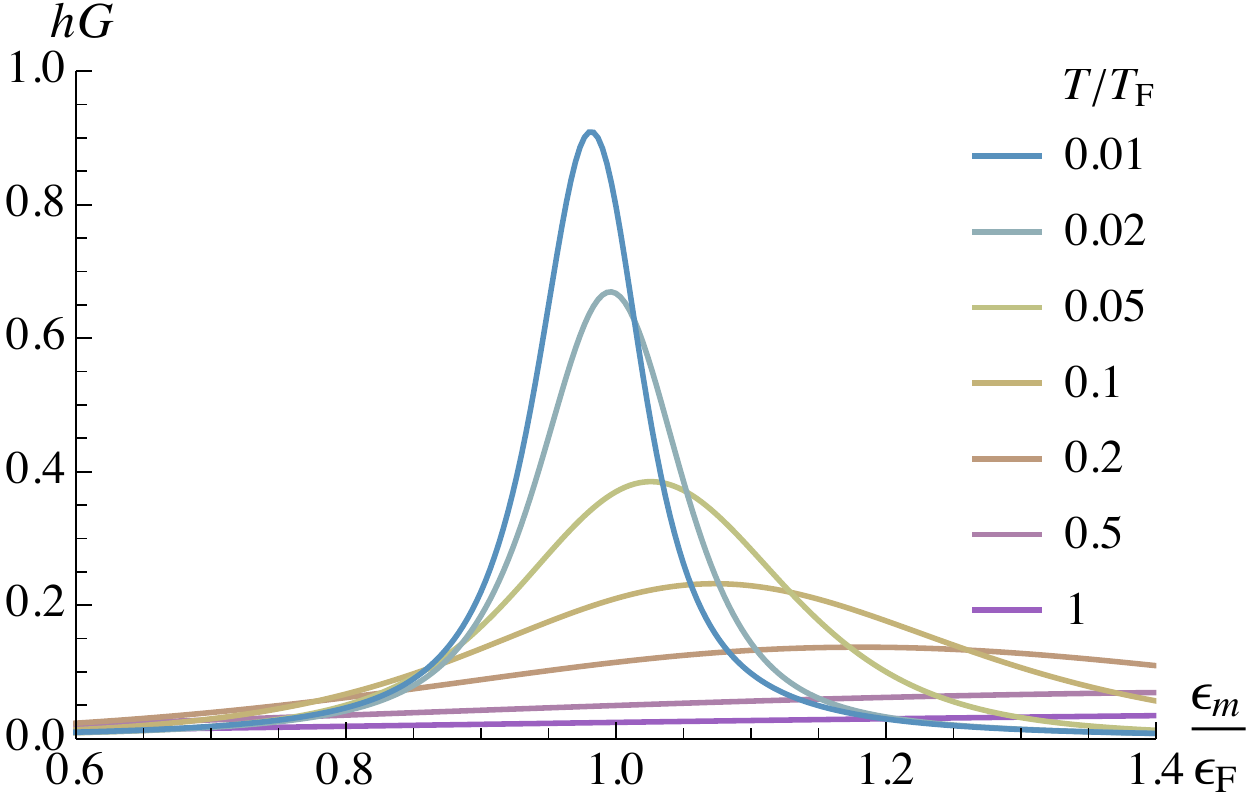}\bigskip\\
 \includegraphics[width=0.92\columnwidth]{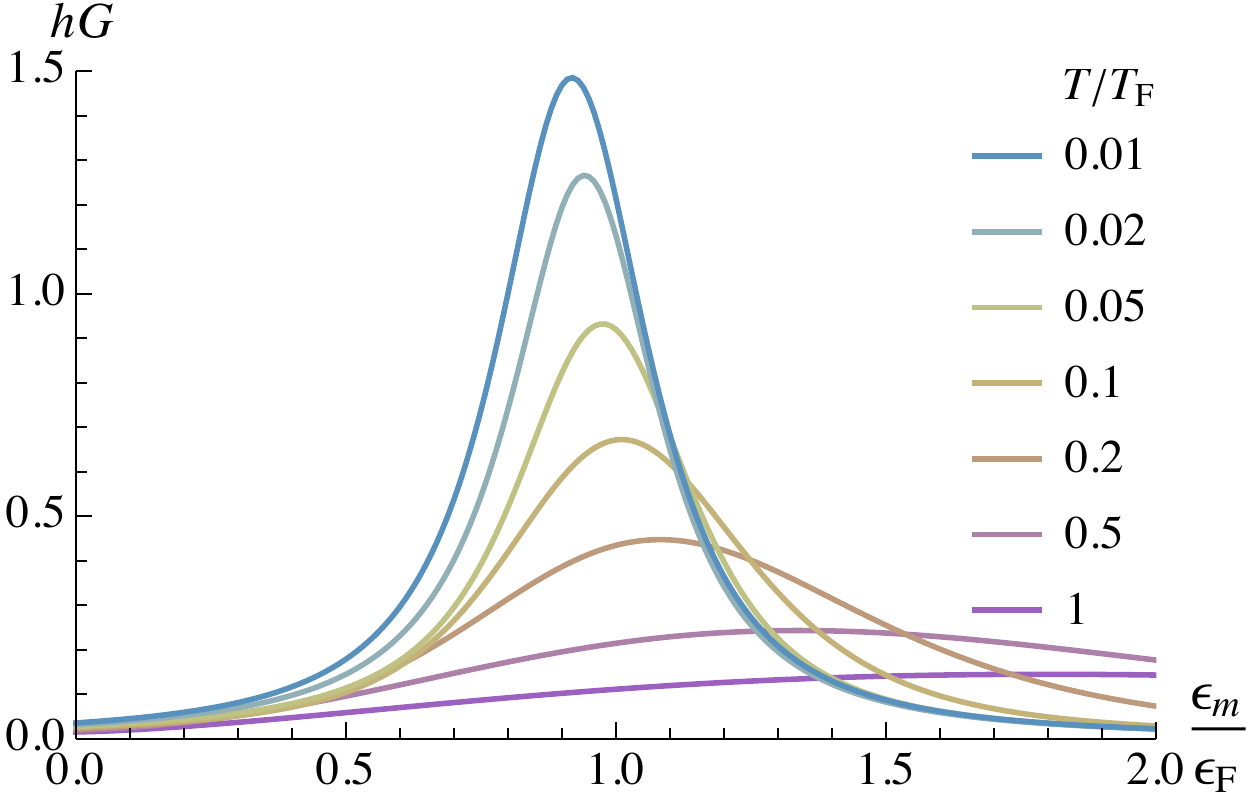}
 \caption{Same as the upper panel of Fig.~\ref{fig:SU3-R1} with $\ell=1$
 and $\Lambda=10\kF$ but for a narrower resonance with $r_m=5\Lambda$
 (upper panel) and in a symmetry broken case to SU(2) with twofold
 degenerate $\eps_m=\eps_{\pm1}$ and decoupled $\eps_0\to\infty$ (lower
 panel).  \label{fig:SU3-R5_SU2-R1}}
\end{figure}

The linear conductance $G$ as a function of the molecular energy is also
computed for different sets of parameters as shown in
Fig.~\ref{fig:SU3-R5_SU2-R1} with $\ell=1$ and $\Lambda=10\kF$ retained.
The upper panel is in the SU(3) symmetric case as before but now for the
larger resonance range of $r_m=5\Lambda$ corresponding to the case where
the confined $B$ atom is heavier than $A$ atoms~\cite{Nishida:2010}.
While the Kondo effect is still observable in the plotted temperature
range, the Kondo temperature gets lower and the peak structure gets
narrower.  On the other hand, the lower panel is aimed at elucidating
the effect of symmetry breaking by setting the resonance range back to
$r_m=\Lambda$.  Here we suppose that the isotropic confinement potential
acting on the impurity $B$ atom is strongly deformed to a uniaxial one
so that the $p$-orbital degeneracy is reduced to twofold with
$\eps_m=\eps_{\pm1}$ and the nondegenerate molecular state is decoupled
with its energy $\eps_0\to\infty$.  The resulting peak structure of the
conductance in the lower panel of Fig.~\ref{fig:SU3-R5_SU2-R1} is found
to remain almost unchanged from that in the upper panel of
Fig.~\ref{fig:SU3-R1} and thus the observability of the Kondo effect is
not impaired by the symmetry breaking from SU(3) to SU(2).

\section{Concluding remarks}
In this Rapid Communication, we proposed and elaborated a simple and
versatile scheme to perform the conductance measurement with ultracold
atoms by employing spin superposition states, which can be implemented,
for example, in a Fermi gas of lithium atoms with impurity ytterbium
atoms~\cite{Hara:2014}.  In particular, we showed that a
confinement-induced $p$-wave or higher partial-wave resonance leads to
the universal logarithmic growth of the conductance toward the low
temperature, which is within reach of observation and thus provides a
clear evidence of the orbital Kondo effect in ultracold atom
experiments.  Not only the proposed transport measurement is applicable
both in and out of equilibrium, but our system is highly tunable and can
be easily extended to a dense Kondo lattice~\cite{Nishida:2013}, which
is difficult to realize in quantum dot systems in spite of its
importance to heavy fermion physics.  It will be interesting to further
incorporate a strong attraction between two components of fermions so
that they form Cooper pairs exhibiting the BCS-BEC crossover.  The
possibility to study such a rich variety of Kondo physics with ultracold
atoms is now opened up.

\acknowledgments
The author thanks Tomosuke Aono, Mikio Eto, Toshimasa Fujisawa, Ryotaro
Inoue, Norio Kawakami, Mikio Kozuma, and Yoshiro Takahashi for valuable
discussions.  This work was supported by JSPS KAKENHI Grant No.~15K17727
and MEXT KAKENHI Grant No.~15H05855.  Part of the numerical calculations
were carried out at the YITP computer facility in Kyoto University.


\begin{thebibliography}{99}

\bibitem{Anderson:1961}
 P.~W.~Anderson,
 ``Localized magnetic states in metals,''
 Phys.\ Rev.\ {\bf 124}, 41-53 (1961).

\bibitem{Hewson:1993}
 A.~C.~Hewson,
 {\em The Kondo Problem to Heavy Fermions\/}
 (Cambridge University Press, Cambridge, UK, 1993).

\bibitem{Kondo:1964}
 J.~Kondo,
 ``Resistance minimum in dilute magnetic alloys,''
 Prog.\ Theor.\ Phys.\ {\bf 32}, 37-49 (1964).

\bibitem{Schrieffer:1966}
 J.~R.~Schrieffer and P.~A.~Wolff,
 ``Relation between the Anderson and Kondo Hamiltonians,''
 Phys.\ Rev.\ {\bf 149}, 491-492 (1966).

\bibitem{Coqblin:1969}
 B.~Coqblin and J.~R.~Schrieffer,
 ``Exchange interaction in alloys with cerium impurities,''
 Phys.\ Rev.\ {\bf 185}, 847-853 (1969).

\bibitem{Coleman:2007}
 P.~Coleman,
 ``Heavy fermions: Electrons at the edge of magnetism,''
 in {\em Handbook of Magnetism and Advanced Magnetic Materials,}
 edited by H.~Kronm\"uller and S.~Parkin
 (John Wiley \& Sons, New York, 2007).

\bibitem{Meir:1991}
 Y.~Meir, N.~S.~Wingreen, and P.~A.~Lee,
 ``Transport through a strongly interacting electron system: Theory of periodic conductance oscillations,''
 Phys.\ Rev.\ Lett.\ {\bf 66}, 3048-3051 (1991).

\bibitem{Meir:1993}
 Y.~Meir, N.~S.~Wingreen, and P.~A.~Lee,
 ``Low-temperature transport through a quantum dot: The Anderson model out of equilibrium,''
 Phys.\ Rev.\ Lett.\ {\bf 70}, 2601-2604 (1993).

\bibitem{Glazman:1988}
 L.~I.~Glazman and M.~E.~Raikh,
 ``Resonant Kondo transparency of a barrier with quasilocal impurity states,''
 JETP Lett.\ {\bf 47}, 452-455 (1988).

\bibitem{Ng:1988}
 T.~K.~Ng and P.~A.~Lee,
 ``On-site Coulomb repulsion and resonant tunneling,''
 Phys.\ Rev.\ Lett.\ {\bf 61}, 1768-1771 (1988).

\bibitem{Grobis:2007}
 M.~Grobis, I.~G.~Rau, R.~M.~Potok, and D.~Goldhaber-Gordon,
 ``The Kondo effect in mesoscopic quantum dots,''
 in {\em Handbook of Magnetism and Advanced Magnetic Materials,}
 edited by H.~Kronm\"uller and S.~Parkin
 (John Wiley \& Sons, New York, 2007).

\bibitem{Wiel:2010}
 W.~G.~van~der~Wiel and S.~De~Franceschi,
 ``Kondo effect in quantum dots,''
 in {\em Handbook of Nanophysics: Nanoparticles and Quantum Dots,}
 edited by K.~D.~Sattler
 (CRC Press, Boca Raton, FL, 2010).

\bibitem{Lewenstein:2012}
 See, for example, M.~Lewenstein, A.~Sanpera, and V.~Ahufinger,
 {\em Ultracold Atoms in Optical Lattices: Simulating Quantum Many-Body Systems\/}
 (Oxford University Press, Oxford, UK, 2012).

\bibitem{Brantut:2012}
 J.-P.~Brantut, J.~Meineke, D.~Stadler, S.~Krinner, and T.~Esslinger,
 ``Conduction of ultracold fermions through a mesoscopic channel,''
 Science {\bf 337}, 1069-1071 (2012).

\bibitem{Stadler:2012}
 D.~Stadler, S.~Krinner, J.~Meineke, J.-P.~Brantut, and T.~Esslinger,
 ``Observing the drop of resistance in the flow of a superfluid Fermi gas,''
 Nature (London) {\bf 491}, 736-739 (2012).

\bibitem{Krinner:2013}
 S.~Krinner, D.~Stadler, J.~Meineke, J.-P.~Brantut, and T.~Esslinger,
 ``Superfluidity with disorder in a thin film of quantum gas,''
 Phys.\ Rev.\ Lett.\ {\bf 110}, 100601 (2013).

\bibitem{Brantut:2013}
 J.-P.~Brantut, C.~Grenier, J.~Meineke, D.~Stadler, S.~Krinner, C.~Kollath, T.~Esslinger, and A.~Georges,
 ``A thermoelectric heat engine with ultracold atoms,''
 Science {\bf 342}, 713-715 (2013).

\bibitem{Krinner:2015}
 S.~Krinner, D.~Stadler, D.~Husmann, J.-P.~Brantut, and T.~Esslinger,
 ``Observation of quantized conductance in neutral matter,''
 Nature (London) {\bf 517}, 64-67 (2015).

\bibitem{Lee:2015}
 J.~Lee, S.~Eckel, F.~Jendrezjewski, C.~J.~Lobb, G.~K.~Campbell, and W.~T.~Hill,~III,
 ``Contact resistance and phase slips in mesoscopic superfluid atom transport,''
 arXiv:1506.08413 [cond-mat.quant-gas].

\bibitem{Husmann:2015}
 D.~Husmann, S.~Uchino, S.~Krinner, M.~Lebrat, T.~Giamarchi, T.~Esslinger, and J.-P.~Brantut,
 ``Connecting strongly correlated superfluids by a quantum point contact,''
 Science {\bf 350}, 1498-1501 (2015).

\bibitem{Kouwenhoven:2001}
 L.~Kouwenhoven and L.~Glazman,
 ``Revival of the Kondo effect,''
 Phys.\ World {\bf 14}, 33-38 (2001).

\bibitem{Nishida:2013}
 Y.~Nishida,
 ``SU(3) orbital Kondo effect with ultracold atoms,''
 Phys.\ Rev.\ Lett.\ {\bf 111}, 135301 (2013).

\bibitem{Recati:2002}
 A.~Recati, P.~O.~Fedichev, W.~Zwerger, J.~von~Delft, and P.~Zoller,
 ``Dissipative spin-boson model and Kondo effect in low dimensional quantum gases,''
 arXiv:cond-mat/0212413.

\bibitem{Falco:2004}
 G.~M.~Falco, R.~A.~Duine, and H.~T.~C.~Stoof,
 ``Molecular Kondo resonance in atomic Fermi gases,''
 Phys.\ Rev.\ Lett.\ {\bf 92}, 140402 (2004).

\bibitem{Duan:2004}
 L.-M.~Duan,
 ``Controlling ultracold atoms in multi-band optical lattices for simulation of Kondo physics,''
 Europhys.\ Lett.\ {\bf 67}, 721-727 (2004).

\bibitem{Paredes:2005}
 B.~Paredes, C.~Tejedor, and J.~I.~Cirac,
 ``Fermionic atoms in optical superlattices,''
 Phys.\ Rev.\ A {\bf 71}, 063608 (2005).

\bibitem{Gorshkov:2010}
 A.~V.~Gorshkov, M.~Hermele, V.~Gurarie, C.~Xu, P.~S.~Julienne, J.~Ye, P.~Zoller, E.~Demler, M.~D.~Lukin, and A.~M.~Rey,
 ``Two-orbital SU($N$) magnetism with ultracold alkaline-earth atoms,''
 Nat.\ Phys.\ {\bf 6}, 289-295 (2010).

\bibitem{Lal:2010}
 S.~Lal, S.~Gopalakrishnan, and P.~M.~Goldbart,
 ``Approaching multichannel Kondo physics using correlated bosons: Quantum phases and how to realize them,''
 Phys.\ Rev.\ B {\bf 81}, 245314 (2010).

\bibitem{Bauer:2013}
 J.~Bauer, C.~Salomon, and E.~Demler,
 ``Realizing a Kondo-correlated state with ultracold atoms,''
 Phys.\ Rev.\ Lett.\ {\bf 111}, 215304 (2013).

\bibitem{Kuzmenko:2015}
 I.~Kuzmenko, T.~Kuzmenko, Y.~Avishai, and K.~Kikoin,
 ``Model for overscreened Kondo effect in ultracold Fermi gas,''
 Phys.\ Rev.\ B {\bf 91}, 165131 (2015).

\bibitem{Sundar:2015}
 B.~Sundar and E.~J.~Mueller,
 ``Proposal to directly observe the Kondo effect through enhanced photo-induced scattering of cold fermionic and bosonic atoms,''
 arXiv:1503.05234 [cond-mat.quant-gas].

\bibitem{Nakagawa:2015}
 M.~Nakagawa and N.~Kawakami,
 ``Laser-induced Kondo effect in ultracold alkaline-earth fermions,''
 Phys.\ Rev.\ Lett.\ {\bf 115}, 165303 (2015).

\bibitem{Patton:2015}
 K.~R.~Patton,
 ``A fully controllable Kondo system: Coupling a flux qubit and an ultracold Fermi gas,''
 arXiv:1508.03016 [cond-mat.quant-gas].

\bibitem{Nishida:2010}
 Y.~Nishida and S.~Tan,
 ``Confinement-induced $p$-wave resonances from $s$-wave interactions,''
 Phys.\ Rev.\ A {\bf 82}, 062713 (2010).

\bibitem{Barnes:1976}
 S.~E.~Barnes,
 ``New method for the Anderson model,''
 J.\ Phys.\ F:\ Met.\ Phys.\ {\bf 6}, 1375-1383 (1976);
%
 ``New method for the Anderson model: II. The $U=0$ limit,''
 J.\ Phys.\ F:\ Met.\ Phys.\ {\bf 7}, 2637-2647 (1977).

\bibitem{Coleman:1984}
 P.~Coleman,
 ``New approach to the mixed-valence problem,''
 Phys.\ Rev.\ B {\bf 29}, 3035-3044 (1984).

\bibitem{Knap:2012}
 M.~Knap, A.~Shashi, Y.~Nishida, A.~Imambekov, D.~A.~Abanin, and E.~Demler,
 ``Time-dependent impurity in ultracold fermions: Orthogonality catastrophe and beyond,''
 Phys.\ Rev.\ X {\bf 2}, 041020 (2012).

\bibitem{transport}
Even when the interaction between $A_\up$ and $A_\down$ atoms is
non-negligible, it does not cause an undesired bulk transport between the
two superposition states, as is evident from its expression of
$\int\!d\r\,\psi_{A\up}^\+(\r)\psi_{A\down}^\+(\r)\psi_{A\down}(\r)\psi_{A\up}(\r)
=\int\!d\r\,\psi_{A+}^\+(\r)\psi_{A-}^\+(\r)\psi_{A-}(\r)\psi_{A+}(\r)$.

\bibitem{Meir:1992}
 Y.~Meir and N.~S.~Wingreen,
 ``Landauer formula for the current through an interacting electron region,''
 Phys.\ Rev.\ Lett.\ {\bf 68}, 2512-2515 (1992).

\bibitem{decoupling}
It is worthwhile to note that $A_\up$ and $A_\down$ atoms are not
decoupled if $\Delta\mu\neq0$ because the chemical potential terms are
expressed as $\sum_{\sigma=\pm}\mu_\pm\psi_{A\sigma}^\+\psi_{A\sigma}
=\mu\,(\psi_{A\up}^\+\psi_{A\up}+\psi_{A\down}^\+\psi_{A\down})
+(\Delta\mu/2)(\psi_{A\up}^\+\psi_{A\down}+\psi_{A\down}^\+\psi_{A\up})$.

\bibitem{Georges:1996}
 A.~Georges, G.~Kotliar, W.~Krauth, and M.~J.~Rozenberg,
 ``Dynamical mean-field theory of strongly correlated fermion systems and the limit of infinite dimensions,''
 Rev.\ Mod.\ Phys.\ {\bf 68}, 13-125 (1996).

\bibitem{Bickers:1987}
 N.~E.~Bickers,
 ``Review of techniques in the large-$N$ expansion for dilute magnetic alloys,''
 Rev.\ Mod.\ Phys.\ {\bf 59}, 845-939 (1987).

\bibitem{Costi:1994}
 T.~A.~Costi, A.~C.~Hewson, and V.~Zlati\'c,
 ``Transport coefficients of the Anderson model via the numerical renormalization group,''
 J.\ Phys.\ Condens.\ Matter {\bf 6}, 2519-2558 (1994).

\bibitem{Goldhaber-Gordon:1998}
 D.~Goldhaber-Gordon, J.~G\"ores, M.~A.~Kastner, H.~Shtrikman, D.~Mahalu, and U.~Meirav,
 ``From the Kondo regime to the mixed-valence regime in a single-electron transistor,''
 Phys.\ Rev.\ Lett.\ {\bf 81}, 5225-5228 (1998).

\bibitem{Carmi:2011}
 A.~Carmi, Y.~Oreg, and M.~Berkooz,
 ``Realization of the SU($N$) Kondo effect in a strong magnetic field,''
 Phys.\ Rev.\ Lett.\ {\bf 106}, 106401 (2011).

\bibitem{Hara:2014}
 H.~Hara, H.~Konishi, S.~Nakajima, Y.~Takasu, and Y.~Takahashi,
 ``A three-dimensional optical lattice of ytterbium and lithium atomic gas mixture,''
 J.\ Phys.\ Soc.\ Jpn.\ {\bf 83}, 014003 (2014).

\end{thebibliography}
\end{document}